\documentclass[12pt]{iopart}

\usepackage{color}
\usepackage{epsfig,amsfonts}

\def\be{\begin{equation}}
\def\ee{\end{equation}}

\def\bea{\begin{eqnarray}}
\def\eea{\end{eqnarray}}

\def\e{\epsilon}

\def\ben{\begin{enumerate}}
\def\een{\end{enumerate}}

\def\bea{\begin{eqnarray}}
\def\eea{\end{eqnarray}}

\begin{document}

\title[Richardson-Gaudin models with one higher spin]{Determinant representation of the domain-wall boundary condition partition function of a Richardson-Gaudin model containing one arbitrary spin}
\author{Alexandre Faribault$^{1}$, Hugo Tschirhart$^{1,2}$ and Nicolas Muller$^{1,3}$}
\address{$^{1}$ Groupe de Physique Statistique, Institut Jean Lamour (CNRS UMR 7198), Universit\'{e} de Lorraine Nancy,
B.P. 70239, FÐ54506 Vandoeuvre-l\`{e}s-Nancy Cedex, France.
\\
$^{2}$ Applied Mathematics Research Center, Coventry University, Coventry, England.
\\
$^{3}$ Universit\'{e} Joseph Fourier, Grenoble, France.}
\ead{alexandre.faribault@univ-lorraine.fr}
\begin{abstract}

In this work we present a determinant expression for the domain-wall boundary condition partition function of rational (XXX) Richardson-Gaudin models which, in addition to $N-1$ spins $\frac{1}{2}$, contains one arbitrarily large spin $S$.

The proposed determinant representation is written in terms of a set of variables which, from previous work, are known to define eigenstates of the quantum integrable models belonging to this class as solutions to quadratic Bethe equations. Such a determinant can be useful numerically since systems of quadratic equations are much simpler to solve than the usual highly non-linear Bethe equations. It can therefore offer significant gains in stability and computation speed. 
\end{abstract}

\date{\today}

\section{Introduction}

The $N \times N$ Cauchy matrix $C^{\left\{\nu_1 \dots \nu_N\right\}}_{\left\{\epsilon_1 \dots \epsilon_N\right\}}$ is defined by matrix elements: 

\bea
C_{ij} = \frac{1}{\nu_i - \epsilon_j}, 
\eea

\noindent built out of two sets $\left\{\nu_1 \dots \nu_N\right\}$ and $\left\{\epsilon_1 \dots \epsilon_N\right\}$ of cardinality $N$ for which every element of both sets is supposed distinct. In \cite{faridet}, it was shown through a recursive proof inspired by \cite{kmt} that its permanent can be written as the determinant of an $N \times N$ matrix:
\bea
\mathrm{Perm} \ C^{\left\{\nu_1 \dots \nu_N\right\}}_{\left\{\epsilon_1 \dots \epsilon_N\right\}} = \mathrm{Det} \ J^{\left\{\nu_1 \dots \nu_N\right\}}_{\left\{\epsilon_1 \dots \epsilon_N\right\}}
\nonumber
 \eea
\noindent defined as
\bea
J_{ij}= 
\left\{
\begin{array}{cc}
\sum_{k\ne i}^N \frac{1}{\epsilon_i - \epsilon_k}  - \sum_{k = 1}^N \frac{1}{\epsilon_i- \nu_k} &   i=j  \\
 \frac{1}{\epsilon_i-\epsilon_j} &  i \ne j
  \end{array}
\right..
\eea

\noindent Know that an alternative proof (De Nardis, J.  {\it private communication}) involves using Borchardt's identity \cite{singer}:

\bea
\mathrm{Det} \ C^{\left\{\nu_1 \dots \nu_N\right\}}_{\left\{\epsilon_1 \dots \epsilon_N\right\}} \  \mathrm{Perm}  \ C^{\left\{\nu_1 \dots \nu_N\right\}}_{\left\{\epsilon_1 \dots \epsilon_N\right\}} = \mathrm{Det} \ M^{\left\{\nu_1 \dots \nu_N\right\}}_{\left\{\epsilon_1 \dots \epsilon_N\right\}}
\eea

 \noindent with $M_{i,j} =\frac{1}{(\nu_i-\epsilon_j)^2}$. The known inverse of the Cauchy matrix then allows one to find the $J^{\left\{\nu_1 \dots \nu_N\right\}}_{\left\{\epsilon_1 \dots \epsilon_N\right\}}$ matrix through the direct calculation of $J = C^{-1} M$.

The Cauchy permanent corresponds to the domain-wall boundary condition partition function of rational (XXX) Richardson-Gaudin quantum integrable model as well as the scalar product between an arbitrary off-the-shell Bethe state and an arbitrary off-the-shell dual Bethe state in these systems. Therefore, the proposed determinant expression finds a direct application for these models and has been used to vastly improve the numerical approaches to their non-equilibirum dynamics \cite{centralfaribault1,centralfaribault2}. It has also recently allowed a similar construction for XXZ-Richardson-Gaudin models for which the partition function can be recast into the permanent of a Cauchy-matrix through the introduction of an arbitrary auxiliary level \cite{XXZ}.

Additionnally, a similar construction, which finds application for spin-boson realisations of the generalized Gaudin-algebra, is based on the mathematical proof found in \cite{bosonhugo} that:

\bea
\sum_{E \in S^{(N)}} \mathrm{Perm} \ C^{\{E_1 \dots E_N\}}_{\{\epsilon_1 \dots \epsilon_N\}} =  \mathrm{Det} \ \tilde{J}^{\{\nu_1 \dots \nu_{N+M}\}}_{\{\epsilon_1 \dots \epsilon_N\}}
\nonumber
\eea

\noindent with
\bea
\tilde{J}_{ij}= 
\left\{
\begin{array}{cc}
\sum_{k\ne i}^N \frac{1}{\epsilon_i - \epsilon_k}  - \sum_{k = 1}^{N+M} \frac{1}{\epsilon_i- \nu_k} &   i=j  \\
 \frac{1}{\epsilon_i-\epsilon_j} &  i \ne j
  \end{array}
\right.
\eea

\noindent where the sum is over every subset of cardinality $N$ one can build out of a set $\{\nu_1 \dots \nu_{N+M}\}$ of larger cardinality $N+M$. Here, $C^{\{E_1 \dots E_N\}}_{\{\epsilon_1 \dots \epsilon_N\}} $ is simply the $N\times N$ Cauchy matrix built out of the fixed set $\{\epsilon_1 \dots \epsilon_N\}$ and one of the $N$-subsets of $\{\nu_1 \dots \nu_{N+M}\}$.

Despite subtle issues in the construction of the dual representation of a given eigenstate \cite{bosonhugo}, the equivalence between this sum of permanents and the much simpler determinant representation gives us again a determinant expression for the overlap of an arbitrary (off-the-shell) Bethe state and an arbitrary dual Bethe state. One should also point out that a completely distinct approach using a pseudo-deformation of the algebra has also recently been used in \cite{bosonbelge} to demonstrate the same results.

In both cases, the main interest  of these determinant expressions is that they are explicitly written in terms of variables which, in the application to their respective Bethe ansatz solvable models, obey quadratic Bethe equations. It therefore becomes numerically much more accessible than the traditional Bethe roots (the rapidities $\nu_i$) which are used in the more traditional Slavnov-Izergin-like determinants \cite{links,slavnov,izerginold,izergin,korepinbook}.

In this work, we build a similar determinant representation for the domain-wall boundary partition function for Richardson-Gaudin models realized in terms of one spin of arbitrary length $S$ and a collection of $N-1$ spins $\frac{1}{2}$.

We show that for an actual set $\{\nu_1 \dots \nu_{2S+N-1}\}$ and a multiset $\{\epsilon_1, \epsilon_1 \dots \epsilon_1, \epsilon_2,\epsilon_3, \epsilon_4 \dots  \epsilon_N\}$ of the same cardinality, the first element $\epsilon_1$ being repeated $2 S$ times, the permanent of the $(2S+N-1)\times (2S+N-1)$ "Cauchy"-matrix also has an $N \times N$ determinant representation expressed in terms of the precise set of variables which obey quadratic Bethe equations for the quantum integrable systems of interest here.  

In the next section, we first present a brief review of the properties of interest for Richardson-Gaudin models built out exclusively of spins $\frac{1}{2}$. Section \ref{basicsofz} then presents the generalization to systems which also contain one higher spin, reviewing the Bethe ansatz and describing the known permanent expression for the domain-wall boundary partition function of interest. Section \ref{variables} then introduces the set of variables our determinant is to be expressed in terms of, while Section \ref{conditions} describes the three necessary and sufficient conditions which any representation of this partition function needs to obey. Finally, Section \ref{detfinal} introduces the proposed determinant and verifies its validity by showing that it does indeed satisfy the needed set of conditions.

\section{XXX Richardson-Gaudin built out of spins-$\frac{1}{2}$}
\label{normal1/2}
Let us first briefly recall the main results from \cite{faridet} which dealt with the specific case of rational (XXX) models built out exclusively from $N$ spins-$\frac{1}{2}$. In this specific case, the realisation of the generalized Gaudin algebra is given by \cite{gaudin, ortiz}:

\bea
S^+(u) &=& \sum_{i=1}^N \frac{S^+_i}{u-\epsilon_i}
\nonumber\\
S^-(u) &=& \sum_{i=1}^N \frac{S^-_i}{u-\epsilon_i}
\nonumber\\
S^z(u) &=&\frac{1}{g} - \displaystyle \sum_{i=1}^N \frac{S^z_i}{u-\epsilon_i}\,
\label{algebra}
\eea

\noindent for which the eigenstates of $S^2(u) \equiv S^z(u)S^z(u)+\frac{1}{2} S^+(u)S^-(u)+\frac{1}{2} S^-(u)S^+(u)$, common to every $S^2(u)\  \forall \ u \in \mathbb{C}$, are of the form:

\bea
\left|\lambda_1 \dots \lambda_{M}\right> \equiv \prod_{i=1}^M S^+(\lambda_i)\left|\downarrow \downarrow \dots \downarrow \right>.
\eea

These generic Bethe states become eigenstates of the "transfer matrix" $S^2(u)$, provided the set of Bethe roots $\left\{\lambda_1 \dots \lambda_M\right\}$ are solutions of a system of $M$ non-linear algebraic equations: the Bethe equations, which are explicitly given (for each $i=1,2 \dots M$) by
\bea
 \frac{2}{g}&=&\sum_{j=1 (\ne i)}^M \frac{2}{\lambda_i-\lambda_j}+\sum_{k=1}^N\frac{1}{\epsilon_k-\lambda_i}.
\label{normalbe}
\eea

One of the main motivations behind this earlier work \cite{faridet} was to build numerically tractable expressions for scalar products and form factors (matrix elements) of local spin operators which would be explicitly given in terms of the $N$ variables: 

\bea
\Lambda_i \equiv \sum_{k=1}^M \frac{1}{\epsilon_i-\lambda_k}.
\eea

This set of variables corresponds to the non-trivial (state-dependent) part of the eigenvalues of the conserved charges of these models. Not only do they have an important physical significance, they also allow one to build an alternative set of Bethe equations which, it turns out, is much simpler than the original ones (\ref{normalbe}). Indeed, one can equivalently define the eigenstates of the system in terms of $\Lambda_i $ provided these $N$ variables are solutions to the $N$ quadratic "Bethe equations" \cite{babelon,baode1}:

\bea
\Lambda_i^2
&=&
 \sum_{j=1 (\ne i)}^N \frac{\Lambda_i-\Lambda_j}{\e_i-\epsilon_j}
+ \frac{2}{g}\Lambda_i.
\label{quadeqs}
\eea

Just like the Heine-Stieltjes approach \cite{heinstil1,heinstil2,heinstil3,heinstil4,heinstil5}, this allows simpler numerical approaches specific to these models.

The basic quantity which was used in \cite{faridet}, to build the relevant $N\times N$ determinant expressions for scalar products and form factors is the domain-wall boundary condition partition function:
\bea
\left<\uparrow \uparrow \dots \uparrow \right| \prod_{i=1}^N S^+(\nu_i)\left| \downarrow \downarrow \dots \downarrow\right> =  \mathrm{Perm} \ C^{\left\{\nu_1 \dots \nu_N\right\}}_{\left\{\epsilon_1 \dots \epsilon_N\right\}} = \mathrm{det}{\tilde{J}}_{N \times N}
\eea
\noindent with
\bea
\tilde{J}_{ij}= 
\left\{
\begin{array}{cc}
\sum_{k\ne i}^N \frac{1}{\epsilon_i - \epsilon_k}  - \sum_{k = 1}^N \frac{1}{\epsilon_i- \nu_k} &   i=j  \\
 \frac{1}{\epsilon_i-\epsilon_j} &  i \ne j
  \end{array}
\right. ,
\eea

\noindent an expression which is valid for arbitrary $\left\{\nu_1 \dots \nu_N\right\} \in \mathbb{C}^N$.

Since it will become the starting point of the recursive proof this work is built on, we point out immediately that an alternative determinant representation can be built by simply changing the signs of every off-diagonal element. Indeed, since transposition leaves a determinant invariant, in this particular case where off-diagonal elements are related by $\tilde{J}_{ij} = -\tilde{J}_{ji}$, one can also write:
\bea
\left<\uparrow \uparrow \dots \uparrow \right| \prod_{i=1}^N S^+(\nu_i)\left| \downarrow \downarrow \dots \downarrow\right> =  \mathrm{det}{J}_{N \times N}
\eea
\noindent with
\bea
J_{ij}= 
\left\{
\begin{array}{cc}
\sum_{k\ne i}^N \frac{1}{\epsilon_i - \epsilon_k}  - \sum_{k = 1}^N \frac{1}{\epsilon_i- \nu_k} &   i=j  \\
 \frac{1}{\epsilon_j-\epsilon_i} &  i \ne j
  \end{array}
\right. .
\label{Jwithsigns}
\eea

Since it is valid for arbitrary values of the $N$ rapidities it is also trivially usable to represent the overlap of an arbitrary off-shell Bethe state $\prod_{i=1}^M S^+(\lambda_i)\left| \downarrow \downarrow \dots \downarrow\right> $ and an arbitrary off-shell dual Bethe state: $\prod_{i=1}^{N-M} S^-(\mu_i) \left|\uparrow \uparrow \dots \uparrow \right> $. Out of the rich dualities \cite{duallinks} present in these models, a simple dual formulation of the Bethe ansatz is easily found by changing the quantization axis from $+\hat{z}$ to $-\hat{z}$. One can then use the inverse quantum scattering method \cite{korepinbook} to find Bethe equations which define eigenstates in both the normal and the dual representation. It turns out that the correspondence between a given normal eigenstate and its dual representation is simple in terms of the eigenvalue-based variables as it is given by the transformation:

\bea
\Lambda^\mu_i = \Lambda^\lambda_i - \frac{2}{g},
\nonumber\\
 \sum_{k=1}^{N-M} \frac{1}{\epsilon_i-\mu_k} = \sum_{k=1}^{M} \frac{1}{\epsilon_i-\lambda_k} - \frac{2}{g}.
\label{trans}
\eea

One should finally know that a generic off-shell Bethe state $\prod_{i=1}^M S^+(\lambda_i)\left| \downarrow \downarrow \dots \downarrow\right> $ with arbitrary $\lambda$s does not have a dual representation, any on-shell state (eigenstate) necessarily does.

\section{Partition function with one arbitrarily large spin}
\label{basicsofz}

Starting from expression (\ref{Jwithsigns}), we will set up a recursive way to build a similar determinant expression for the case where one of the spins (without loss of generality we systematically choose $S_1$) is raised from a $S=\frac{1}{2}$ to $S=1$ to $S=\frac{3}{2}$ and so on, up to an arbitrary $S=\frac{d}{2}$.

Integrability and the Bethe ansatz solution of this particular system do not rely on the representation of the spin, be it spin $1/2$ or higher. In every case, the eigenstates are still built out of the same operator $S^+(\lambda)$ defined in eq. (\ref{algebra}) acting on the fully down polarized state, i.e. eigenstate of every $S^z_i$ with the lowest possible (negative) eigenvalues $m_z$. The dual is naturally built using $S^-(\lambda)$ acting on the fully up polarized state. As was explicitly shown in \cite{baode2}, the Bethe equations whose solutions define eigenstates $\prod_{i=1}^M S^+(\lambda) \left|\Downarrow_S, \downarrow,\downarrow, \dots \downarrow\right>$ of these systems, can also be recast into a set of quadratic equations which explicitly depend on the set of variables $\{\Lambda(\epsilon_1),\Lambda^{(1)}(\epsilon_1) \dots \Lambda^{(2S)}(\epsilon_1),\ \Lambda(\epsilon_2),\Lambda(\epsilon_3) \dots \ \Lambda(\epsilon_N)  \}$, where $\Lambda(z) = \sum_{i=1}^M \frac{1}{z-\lambda_i}$ and $\Lambda^{(n)}(z) \equiv \frac{\partial^n\Lambda(z) }{\partial z^n}$.

Since the first spin can now accommodate more than a single excitation ($S^+_1S^+_1 =0$ only for spins 1/2), going from the fully down to the fully up polarized state now requires at total of $\Omega = 2S + (N-1)$ excitations. The generic expression for the domain-wall boundary partition function of  interest for such a system is still given by a permanent of dimension $\Omega \times \Omega$:
\bea
\left<  \Uparrow^S_{1} \dots  \uparrow_{N}\right|S^+(\nu_1) S^+(\nu_2) \dots S^+(\nu_\Omega) \left| \Downarrow^S_{1} \dots \downarrow_{N}  \right> = \mathrm{Perm} \ \tilde{C}_{\Omega \times \Omega}.
\label{zdefinition}
\eea

\noindent Here $ \tilde{C}$ is the "Cauchy"-matrix built out of the set $\{\nu_1 \dots \nu_\Omega\}$ while, this time, the second "set" actually becomes the multiset $\{\epsilon_1, \epsilon_1 \dots \epsilon_1, \epsilon_2,\epsilon_3, \epsilon_4 \dots  \epsilon_N\}$ with the first element $\epsilon_1$ repeated $2 S$ times. That is to say that $2S$ of the rapidities have to be associated with $\epsilon_1$ and the remaining ones are each associated to one of the other spin's $\epsilon_i$. The resulting product of the terms $\frac{1}{\nu_i-\epsilon_j}$ is then summed over possible mappings.

It also remains true that inverting the quantization axis guarantees the existence of a dual Bethe ansatz so that the partition function (\ref{zdefinition}) we are interested in corresponds to the scalar product of an arbitrary off-shell Bethe state $\prod_{i=1}^{M} S^+(\lambda_i) \left|\Downarrow_S \ \downarrow \dots \downarrow \right> $ and an arbitrary off-shell dual Bethe state $\prod_{i=1}^{\Omega-M} S^-(\mu_i) \left|\Uparrow_S \ \uparrow \dots \uparrow \right> $.

For definiteness, the simplest such scenario is the case of a single spin $1$ and a single spin $\frac{1}{2}$ for which the partition function of interest can be explicitly written as:

\bea
&& Z^{S; \left( \frac{1}{2}\right)}_{\left\{\nu_1, \nu_2 ,\nu_{3}\right\}} = \left<  \Uparrow_1\uparrow_2\right|S^+(\nu_1) S^+(\nu_2) S^+(\nu_3)\left| \Downarrow_1\downarrow_2\right> \nonumber\\ &&= \frac{1}{(\nu_1 - \epsilon_1)(\nu_2 - \epsilon_1)(\nu_3-\epsilon_2)}  +  \frac{1}{(\nu_1 - \epsilon_1)(\nu_3 - \epsilon_1)(\nu_2-\epsilon_2)} \nonumber\\ &&+ \frac{1}{(\nu_2 - \epsilon_1)(\nu_1 - \epsilon_1)(\nu_3 - \epsilon_2)}+ \frac{1}{(\nu_2 - \epsilon_1)(\nu_3 - \epsilon_1)(\nu_1 - \epsilon_2)}\nonumber\\ && +  \frac{1}{(\nu_3 - \epsilon_1)(\nu_1 - \epsilon_1)(\nu_2-\epsilon_2)} + \frac{1}{(\nu_3 - \epsilon_1)(\nu_2 - \epsilon_1)(\nu_1 - \epsilon_2)}\nonumber\\ && = \frac{2}{(\nu_1 - \epsilon_1)(\nu_2 - \epsilon_1)(\nu_3-\epsilon_2)}+\frac{2}{(\nu_2 - \epsilon_1)(\nu_3-\epsilon_1)(\nu_1 - \epsilon_2)}\nonumber\\ && +\frac{2}{(\nu_1 - \epsilon_1)(\nu_3-\epsilon_1)(\nu_2 - \epsilon_2)}. 
\eea

In general one has

\bea
Z^{S; \left( \frac{1}{2}\right)^{\otimes_{N-1}}}_{\left\{\nu_1 \dots \nu_{\Omega}\right\}} =  \sum_{A \in {R^{(2S)}}}\sum_{B \in B^{(\tilde{A})}} \frac{(2S)! }{\prod_{i=1}^{2S}\left(A_i-\epsilon_1\right)  }\frac{1}{\prod_{i=1}^{N-1} \left(B_i - \epsilon_{i+1}\right)}
\label{genz}
\eea
 
 \noindent where $R^{(2S)}$ is the set composed of every subset of $\{\nu_1 \dots \nu_\Omega\}$  with given cardinality $2S$ and $B^{(\tilde{A})} = \{ (R\backslash A) \}$ is the set of all $(N-1)$-tuples (permutations) one can build out of the elements of the relative complement $ R\backslash A$, therefore excluding any rapidity already present in $A$. Having chosen a given set of $2S$ rapidities to associate with the first spin ($\epsilon_1$), we then sum over all bijections between the $N-1$ remaining rapidities and the $N-1$ inhomogeneity parameters associated with the spins $\frac{1}{2}$. Summing these contributions gives us the desired partition function which also corresponds to the "partially-homogeneous" limit obtained from having $2S+N-1$ spins-$\frac{1}{2}$ of which the first $2S$ share the same inhomogeneity parameter $\epsilon_1$.

This explicit construction retains the basic features of the Cauchy permanent obtained without repeated $\epsilon_1$, in that it has exclusively single poles in each of the $\nu_i$ variables. The residue at any of these poles will reproduce the exact same permanent structure in terms of the set and multiset from which $\nu_i$ and one instance of $\epsilon_j$ have respectively been removed. Moreover, since each rapidity $\nu_i$ necessarily appears in every term of the permanent,  it is obvious that, in the limit when any $\nu_i\to \infty$, the partition function tends to zero.

\section{Set of variables}
  \label{variables}

Considering that, in each $\nu_i$  variable, only single poles can appear in the partition function it can therefore be written in a way which explicitly depends on combinations which reproduce this structure. Starting from the $Q(z) = \prod_{i=1}^\Omega (z-\nu_i)$ polynomial, one can build a hierarchy of such rational functions:

\bea
\Gamma_0(z)& \equiv & - \frac{Q(z)}{Q(z)}  = - 1
\nonumber\\
\Gamma_1(z)& \equiv &- \frac{Q'(z)}{Q(z)}  = - \sum_{i=1}^\Omega \frac{1}{z-\nu_i} = -\Lambda(z)
\nonumber\\
\Gamma_2(z)& \equiv &- \frac{Q''(z)}{Q(z)}  = - \sum_{i_1\ne i_2}^\Omega \frac{1}{(z-\nu_{i_1})(z-\nu_{i_2})} = - \Lambda'(z) -  \Lambda(z)^2
\nonumber\\
\Gamma_3(z)& \equiv & - \frac{Q'''(z)}{Q(z)}  = - \sum_{i_1\ne i_2\ne i_3}^\Omega \frac{1}{(z-\nu_{i_1})(z-\nu_{i_2})(z-\nu_{i_3})} = - \Lambda''(z) - 3 \Lambda(z) \Lambda'(z)-  \Lambda^3(z)
\nonumber\\
&&\vdots
\eea

This set of $\Gamma_i(z)$ functions can therefore be defined recursively by noticing that, taking the derivative of $\Gamma_{n-1}(z)$, one finds:

\bea
\frac{\partial}{\partial z}\Gamma_{n-1}(z) =-  \frac{Q^{(n)}(z)}{Q(z)} +\frac{Q^{(n-1)}(z)}{Q(z)^2}Q'(z) = \Gamma_{n}(z) - \Lambda(z) \Gamma_{n-1}(z).
\nonumber\\
\Gamma_n(z) = \frac{\partial}{\partial z}\Gamma_{n-1}(z) + \Lambda(z) \Gamma_{n-1}(z).
\label{recursivevariables}
\eea

In terms of the set of $\Lambda^{(a)}(z)$ (the $a^{\mathrm{th}}$ derivative of $\Lambda(z)$), one can explicitly verify that the solution to this recurrence is given by:

\bea
\Gamma_n(z) = \sum_{\{k_0, k_1 \dots  \ k_n \} }C^n_{\{k_0, k_1 \dots k_n \} } \left[ \prod_{a=0}^n\left(\Lambda^{(a)}(z)\right)^{k_a}\right] ,
\eea 

\noindent with
\bea
C^n_{\{k_0, k_1 \dots  k_n \} }
= \frac{n !} {\prod_{a=0}^n \left[\left[(a+1)!\right]^{k_a}k_a!\right]} \delta_{\left[\sum_{a=0}^n(a + 1) k_a\right], n} \ .
\label{coefficients}
\eea

Supposing the form valid for $n-1$, we find that recursion (\ref{recursivevariables}) will then be verified since:
\bea
&&\frac{\partial}{\partial z}\Gamma_{n-1}(z) + \Lambda(z) \Gamma_{n-1}(z) = 
\frac{\partial\left(\displaystyle \sum_{\{k_0, k_1 \dots  \ k_{n} \} }C^{n-1}_{\{k_0, k_1 \dots k_{n } \} } \left[ \prod_{a=0}^n\left(\Lambda^{(a)}(z)\right)^{k_a}\right]\right)}{\partial z}\nonumber\\  &&+ \Lambda^{(0)}(z) \left( \sum_{\{k_0, k_1 \dots  \ k_{n} \} }C^{n-1}_{\{k_0, k_1 \dots k_{n } \} } \left[ \prod_{a=0}^n\left(\Lambda^{(a)}(z)\right)^{k_a}\right]\right)\nonumber\\ &=&
\displaystyle \sum_{b=0}^{n-1} \sum_{\{k_0, k_1 \dots  \ k_{n} \} } k_bC^{n-1}_{\{k_0, k_1 \dots k_{n } \} } \left[ \prod_{a=0 (\ne b)}^{n}\left(\Lambda^{(a)}(z)\right)^{k_a}\right]\left(\Lambda^{(b)}(z)\right)^{k_b-1}\left(\Lambda^{(b+1)}(z)\right)^{k_{b+1}+1}\nonumber\\  &&+\left( \sum_{\{k_0, k_1 \dots  \ k_{n} \} }C^{n-1}_{\{k_0, k_1 \dots k_{n } \} } \left[ \left(\Lambda^{(0)}(z)\right)^{k_0+1}\prod_{a=1}^n\left(\Lambda^{(a)}(z)\right)^{k_a}\right]\right),
\eea

\noindent which, regrouping the terms with a given set of powers $\left(k_0,k_1 \dots k_n \right)$, can be rewritten as:
\bea
&&\frac{\partial}{\partial z}\Gamma_{n-1}(z) + \Lambda(z) \Gamma_{n-1}(z) =  \sum_{\{k_0, k_1 \dots  \ k_{n} \} }\tilde{C}^{n}_{\{k_0, k_1 \dots k_{n } \} } \prod_{a=0}^n\left(\Lambda^{(a)}(z)\right)^{k_a}
\eea
\noindent with coefficients 
\bea
\tilde{C}^n_{\{k_0, k_1 \dots  k_n \} } = C^{n-1}_{\{k_0 -1 , k_1\dots k_{n } \} }+ \sum_{b=0}^{n-1} k_b C^{n-1}_{\{k_0, k_1 \dots k_b + 1, k_{b+1} - 1\dots k_{n } \} }. 
\eea

It is then simple to verify that this last relation is indeed verified by the coefficients proposed in (\ref{coefficients}) since every term respects the $\sum_{a=0}^n (a+1) k_a = n$ condition so that the right hand side can be written as:

\bea
&&C^{n-1}_{\{k_0 -1 , k_1\dots k_{n } \} }+ \sum_{b=0}^{n-1} k_b C^{n-1}_{\{k_0, k_1 \dots k_b + 1, k_{b+1} - 1\dots k_{n } \} } \nonumber\\&&= 
 \left[\frac{(n-1) ! \ (1!) k_0} {\prod_{a=0}^n \left[\left[(a+1)!\right]^{k_a}k_a!\right]}  + \sum_{b=0}^{n-1} k_b\frac{(b+2)! \ k_{b+1}}{(b+1)! \ k_{b}} \frac{(n-1) ! } {\prod_{a=0}^n \left[\left[(a+1)!\right]^{k_a}k_a!\right]}\right] \delta_{\left[\sum_{a=0}^n(a + 1) k_a\right], n} 
 \nonumber\\&&= 
 \left[\frac{(n-1) ! \left[(0+1) k_0+ \sum_{b=0}^{n-1} (b+2) \ k_{b+1}\right]} {\prod_{a=0}^n \left[\left[(a+1)!\right]^{k_a}k_a!\right]}  \right] \delta_{\left[\sum_{a=0}^n(a + 1) k_a\right], n}  \nonumber\\ &&= \left[\frac{ (n-1)! \left[\sum_{b=0}^{n} (b+1) \ k_{b}\right]} {\prod_{a=0}^n \left[\left[(a+1)!\right]^{k_a}k_a!\right]}  \right] \delta_{\left[\sum_{a=0}^n(a + 1) k_a\right], n} = \left[\frac{ (n)! \delta_{\left[\sum_{a=0}^n(a + 1) k_a\right], n} } {\prod_{a=0}^n \left[\left[(a+1)!\right]^{k_a}k_a!\right]}  \right] = C^n_{\{k_0, k_1 \dots  k_n \} }. \nonumber\\
\eea

Verifying the validity of (\ref{coefficients}) for $n=1$ is simple since only $k_0=1$ with $k_{i\ne 0} = 0$ respects the condition imposed by the Kronecker delta, namely that $\sum_{a=0}^n (a+1) k_a = n$. This verification therefore completes the proof.

For the system of interest here, we will build the partition function explicitly in terms of the variables 
$\{\Gamma_1(\epsilon_1) \dots \Gamma_{2S}(\epsilon_1), \Gamma_1(\epsilon_2),\Gamma_1(\epsilon_3) \dots \Gamma_{1}(\epsilon_N)\}$, which as we have shown can themselves be simply built out of $\{\Lambda(\epsilon_1) \dots \Lambda^{(2S)}(\epsilon_1),\ \Lambda(\epsilon_2),\Lambda(\epsilon_3) \dots \ \Lambda(\epsilon_N)  \}$, i.e. on every $\Lambda(\epsilon_i)$ and on the $2S$ first derivatives of $\Lambda(z)$ evaluated at $\epsilon_1$. These are precisely the variables in terms of which the set of quadratic Bethe equations is built and which allow one to build a simpler numerical approach to the problem of finding eigenstates of the system. The proposed determinant would allow us to take full advantage of these simplifications.

\section{Higher spin partition function}
 \label{conditions}

As we mentioned before, the explicit expression for the domain-wall boundary partition function given in (\ref{genz}) is a rational function which contains only single poles for each of the rapidities $\nu_i$ and is fully symmetric under exchange of any two of these parameters. 

It obeys a set of recursive relations linking the partition functions for a variety of systems. Explicitly regrouping the terms where $\nu_{i}$ (any of them, by symmetry) is paired to $\epsilon_1$, one can write it as a sum over the similar partition functions one obtains when the first spin goes from $S \to S - \frac{1}{2}$, after excluding rapidity $\nu_i$:

\bea
Z^{S; \left( \frac{1}{2}\right)^{\otimes_{N-1}}}_{\left\{\nu_1 \dots \nu_{\Omega}\right\}}  = \sum_{i=1}^\Omega \frac{2S}{\nu_i - \epsilon_1} Z^{S-\frac{1}{2} ; \left( \frac{1}{2}\right)^{\otimes_{N-1}}}_{\left\{\nu_1 \dots \nu_{i-1}, \nu_{i+1} \dots \nu_{\Omega}\right\}}.
\label{rele1}
\eea

Identically, one can also write it in terms of the partition functions obtained by excluding any one of the spins $\frac{1}{2}$ (say spin $j$):

\bea
Z^{S ; \left( \frac{1}{2}\right)^{\otimes_{N-1}}}_{\left\{\nu_1 \dots \nu_{\Omega}\right\}}  = \sum_{i=1}^\Omega \frac{1}{\nu_i - \epsilon_j}  Z^{S ; \left( \frac{1}{2}\right)_{\hat{j}}^{\otimes_{N-2}}}_{\left\{\nu_1 \dots \nu_{i-1}, \nu_{i+1} \dots \nu_{\Omega}\right\}}.
\label{relej}
\eea

As we also pointed out, the construction is such that $\lim_{\nu_i \to \infty}Z^{S \left( \frac{1}{2}\right)^{\otimes_{N-1}}}_{\left\{\nu_1 \dots \nu_{\Omega}\right\}} =0$, for any of the rapidities.

These properties can be used to set up a recursive proof for any proposed form, whose starting point will be the previously found representation (\ref{Jwithsigns}) for a collection of $N$ spins $\frac{1}{2}$. To prove the equality of two rational functions (in this case containing only single poles), one simply needs to show that they share the same poles, the same residues at these poles and the same limit at infinity. Thus, any proposed representation for the partition functions can be shown to be valid by simply verifying that these conditions are met for the proposed representation.

Three necessary and sufficient conditions therefore need to be fulfilled in order to validate an expression for the partition function obtained after raising the first spin from a spin $S-\frac{1}{2}$ to a spin $S$ while adding a new rapidity $\nu_{\Omega}$. 

First, from (\ref{rele1}), the residue of this partition function at $\nu_{\Omega} = \epsilon_1$ has to be given by 
\bea
\mathrm{Res}_{\nu_{\Omega} = \epsilon_1}Z^{S; \left( \frac{1}{2}\right)^{\otimes_{N-1}}}_{\left\{\nu_1 \dots \nu_{\Omega}\right\}} = 2S \ Z^{S-\frac{1}{2}; \left( \frac{1}{2}\right)^{\otimes_{N-1}}}_{\left\{\nu_1 \dots \nu_{\Omega-1}\right\}}, 
\label{rese1}
\eea

\noindent which considering the explicit symmetry under exchange of any two rapidities would also be valid for the poles at an arbitrary $\nu_i = \epsilon_1$. This symmetry is guaranteed since the proposed determinant representation will be expressed exclusively in terms of the $\Gamma$ variables, themselves symmetrical under such an exchange.

Secondly, from (\ref{relej}), one needs the residue at  $\nu_{\Omega+1} = \epsilon_N$ to be given by the determinant obtained after removing the last spin:
\bea
\mathrm{Res}_{\nu_{\Omega+1} = \epsilon_N}Z^{S; \left( \frac{1}{2}\right)^{\otimes_{N-1}}}_{\left\{\nu_1 \dots \nu_{\Omega+1}\right\}} = Z^{S; \left( \frac{1}{2}\right)_{\hat{N}}^{\otimes_{N-2}}}_{\left\{\nu_1 \dots \nu_{\Omega}\right\}}, 
\label{resej}
\eea

\noindent again a fact which remains valid for the residue at  $\nu_i = \epsilon_N$ for any of the rapidities. Provided the proposed form is explicitly symmetric under the exchange of any two of the spins-$\frac{1}{2}$, this last condition also immediately leads to a similar result for the residues at  each $\nu_{i} = \epsilon_j$ for any $i \in \{1, \dots \Omega\}$ and $j \in \{2, 3 \dots N\}$.

Any representation which verifies these conditions will therefore have the correct poles and residues and its validity will then only require that the limit at any $\nu_i \to \infty$ be 0. Since our determinant representation will be expressed in terms of the $\Gamma$ variables, it is not only symmetric under exchange of two rapidities but it is also obviously non-diverging for any $\nu_i \to \infty$. These facts imply that, once every pole and residue have been checked to be the right ones, the only possible difference between the known expression and the proposed one could be the addition of a simple constant. It will therefore be sufficient to check that, when every rapidity is taken to infinity, the limit does indeed go to zero. 

The proposed form will consequently be equal to the known permanent form discussed in section \ref{basicsofz} provided we verify the three conditions described in this section: residues at $\epsilon_1$, residues at $\epsilon_{i\ne 1}$ and the limit at $\infty$.

\section{Determinant representation}
\label{detfinal}
  
In the same spirit as the partition function for a collection of spins $\frac{1}{2}$, we construct a $N \times N$ determinant representation such that for every $\nu_i$, the poles at $\nu_i=\epsilon_j$ exclusively appear on line $j$ of the matrix. Moreover, we posit that the diagonal element $J^{S}_{11}$ contains every allowed $\Gamma_{n}(\epsilon_1)$ ($n\in\{0,1 \dots 2S \}$) while off-diagonal elements in the first line do not contain the last one: $\Gamma_{2S}(\epsilon_1)$. The same remains true on the other lines corresponding to a spin-$\frac{1}{2}$ and we therefore have a generic form:

\bea
J^{S}_{11} =(2S)! \sum_{n=0}^{2S} {}^{S}C^{n}_{11} \Gamma_n(\epsilon_1)
\nonumber\\
J^{S}_{1j} =(2S)! \sum_{n=0}^{2S-1} {}^{S}C^{n}_{1j} \Gamma_n(\epsilon_1) \ \ &\forall& \ j \ne 1
\nonumber\\
J^{S}_{ii} = {}^{S}C^{0}_{ii} + {}^{S}C^{1}_{ii} \Gamma_1(\epsilon_i)   \ \ &\forall &\ i \ne 1
\nonumber\\
J^{S}_{ij} = {}^{S}C^{0}_{ij}  \ \ &\forall& \ i \ne 1\ ,\ j \ne i.
\eea

Exchanging both the rows and the columns associated with any two spins $j , j' $ leaves a determinant invariant. Consequently, this makes, as we required, the expression explicitly symmetric under the exchange of two spins. Once again, being built out of the symmetric variables $\Gamma_n(\epsilon_j)$ it is also symmetric under the exchange of any two rapidities $\nu_j, \nu_{j'}$, making both fundamental assumptions of the preceding section valid.

As we will demonstrate, the following set of coefficients ${}^{S}C^{a}_{ij}$ gives a correct representation of the partition function $Z^{S ; \left( \frac{1}{2}\right)^{\otimes_{N-1}}}_{\left\{\nu_1 \dots \nu_{\Omega}\right\}} = \mathrm{det}_{N\times N} J^{S}$
\noindent where
\bea
J^{S}_{ij}= 
\left\{
\begin{array}{cc}
(2S)!\displaystyle\sum_{n=0}^{2S}\left[\displaystyle\sum_{E\in S^{(2S-n)}} \frac{1}{\prod_{k=1}^{2S-n}(\epsilon_1 - E_k)}\right]\Gamma_n(\epsilon_1)   &   i=j=1  \\
-(2S)!\displaystyle\sum_{n=0}^{2S-1}\left[\displaystyle\sum_{p=n}^{2S-1}\displaystyle\sum_{E \in S^{(p-n)}_{\hat{j}}} \frac{2S-p}{(\epsilon_{1}- \epsilon_{j})^{2S-p}}\frac{1}{\prod_{k = 1}^{p-n}(\epsilon_{1} - E_{k})}\right]\Gamma_n(\epsilon_1)  & i=1 \ne j\\
\left[ \frac{2S}{\epsilon_i-\epsilon_1} + \displaystyle\sum_{k \ne i \ne 1}^{N}\frac{1}{\epsilon_i-\epsilon_k} \right] + \Gamma_1(\epsilon_i) &  i = j \ne 1\\
 \frac{1}{\epsilon_j-\epsilon_i} &  i \ne j \ne 1
  \end{array}
\right. ,
\nonumber\\
\label{finalJ}
\eea

\noindent with $S^{(n)}$ the ensemble of multisets built by picking $n$ elements in $\{ \epsilon_2,  \dots \epsilon_N \}$, while $S^{(n)}_{\hat{j}}$ is the ensemble of multisets built by picking $n$ elements in $\{ \epsilon_2,  \dots \epsilon_{j-1}, \epsilon_{j+1}\dots \epsilon_N \}$ (excluding $\epsilon_j$).%

That is to say that on the first line, $C^{n}_{11}$ is, for any order $n$, given by the sum over every possible choice, with repetitions allowed, of $n$ elements out of $\{ \epsilon_2,  \dots \epsilon_N \}$. The off-diagonal elements coefficients are built in the same fashion, except that in column $j$, $\epsilon_j$ has to be picked at least once and the terms are weighted by $2S-p$, the number of times $\epsilon_j$ actually appears.

For example, when combining a spin $\frac{3}{2}$ and two spins $\frac{1}{2}$, one would find the following set of coefficients:

\bea
{}^{\frac{3}{2}}C^{0}_{11} &=& \frac{1}{(\epsilon_{1}-\epsilon_{2})^3}+\frac{1}{(\epsilon_{1}-\epsilon_{3})^3}+\frac{1}{(\epsilon_{1}-\epsilon_{2})^2(\epsilon_{1}-\epsilon_{3})}+\frac{1}{(\epsilon_{1}-\epsilon_{2})(\epsilon_{1}-\epsilon_{3})^2}
\nonumber\\
{}^{\frac{3}{2}}C^{1}_{11} &=& \frac{1}{(\epsilon_{1}-\epsilon_{2})^2}+\frac{1}{(\epsilon_{1}-\epsilon_{3})^2}+\frac{1}{(\epsilon_{1}-\epsilon_{2})(\epsilon_{1}-\epsilon_{3})}
\nonumber\\
{}^{\frac{3}{2}}C^{2}_{11} &=& \frac{1}{(\epsilon_{1}-\epsilon_{2})}+\frac{1}{(\epsilon_{1}-\epsilon_{3})}
\nonumber\\
{}^{\frac{3}{2}}C^{3}_{11} &=& 1
\nonumber\\
{}^{\frac{3}{2}}C^{0}_{12} &=&  \frac{1}{(\epsilon_{2}-\epsilon_{1})}\left(\frac{3}{(\epsilon_{1}-\epsilon_{2})^2}+\frac{1}{(\epsilon_{1}-\epsilon_{3})^2}+\frac{2}{(\epsilon_{1}-\epsilon_{2})(\epsilon_{1}-\epsilon_{3})}\right)
\nonumber\\
{}^{\frac{3}{2}}C^{1}_{12} &=&  \frac{1}{(\epsilon_{2}-\epsilon_{1})}\left(\frac{2}{(\epsilon_{1}-\epsilon_{2})}+\frac{1}{(\epsilon_{1}-\epsilon_{3})}\right)
\nonumber\\
{}^{\frac{3}{2}}C^{2}_{12} &=&  \frac{1}{(\epsilon_{2}-\epsilon_{1})}
\nonumber\\
{}^{\frac{3}{2}}C^{0}_{22} &=&\frac{3}{(\epsilon_{2}-\epsilon_{1})}+\frac{1}{(\epsilon_{2}-\epsilon_{3})}
\nonumber\\
{}^{\frac{3}{2}}C^{1}_{22} &=&1
\nonumber\\
{}^{\frac{3}{2}}C^{0}_{ij} &=&\frac{1}{(\epsilon_{j}-\epsilon_{i})}
\eea

\noindent where one can write the coefficients of the third column by exchanging the role of $\epsilon_2$ with $\epsilon_3$.

It is a simple enough task to verify the residue at $\epsilon_1$ since these poles exclusively appear on the first line of the matrix, from which every term in the determinant can only contain a single one of the first line's element . The residue of the determinant is therefore given as a determinant as well, built form the residues of the matrix elements of the first line while the other matrix elements are found by computing the $\nu_{\Omega} \to \epsilon_1$ limit. Knowing that:

\bea
\lim_{\nu_{\Omega} \to \epsilon_1} \Gamma^{\{\nu_1 \dots \nu_\Omega\}}_n(\epsilon_j) = \Gamma^{\{\nu_1 \dots \nu_{\Omega-1}\}}_n(\epsilon_j) + \frac{1}{\epsilon_1-\epsilon_j} \Gamma^{\{\nu_1 \dots \nu_{\Omega-1}\}}_{n-1}(\epsilon_j) \ \forall  j \ne 1 
\nonumber\\ 
\mathrm{Res}_{\nu_{\Omega} = \epsilon_1} \Gamma^{\{\nu_1 \dots \nu_{\Omega}\}}_n(\epsilon_1) = \Gamma^{\{\nu_1 \dots \nu_{\Omega-1}\}}_{n-1}(\epsilon_1) , 
\eea
\noindent we find:
\bea
\mathrm{Res}_{\nu_{\Omega} = \epsilon_1}\det_{N\times N} J^S = \det_{N\times N} \tilde{J}^{S-\frac{1}{2}} 
\eea
\noindent where $\tilde{J}^{S-\frac{1}{2}} $ is easily shown to correspond to the $J^{S-\frac{1}{2}}$ defined in (\ref{finalJ}) for a spin $S-\frac{1}{2}$:
\bea
\tilde{J}^{S-\frac{1}{2}} _{11} =(2S)! \sum_{n=1}^{2S}  {}^{S}C^{n}_{11} \Gamma^{\{\nu_1 \dots \nu_{\Omega-1}\}}_{n-1}(\epsilon_1) \nonumber\\ \ \  \ \ =2S \ (2S-1)! \sum_{n=0}^{2S-1}  {}^{S}C^{n+1}_{11} \Gamma^{\{\nu_1 \dots \nu_{\Omega-1}\}}_{n}(\epsilon_1) = 2SJ^{S-\frac{1}{2}} _{11} &&
\nonumber\\ 
\tilde{J}^{S-\frac{1}{2}} _{1j} = (2S)!\sum_{n=1}^{2S-1}  {}^{S}C^{n}_{1j} \Gamma^{\{\nu_1 \dots \nu_{\Omega-1}\}}_{n-1}(\epsilon_1) \nonumber\\ \ \ \ \ =2S \ (2S-1)! \sum_{n=0}^{2S-2}  {}^{S}C^{n+1}_{1j} \Gamma^{\{\nu_1 \dots \nu_{\Omega-1}\}}_{n}(\epsilon_1)= 2SJ^{S-\frac{1}{2}} _{1j} \ \ \ &\forall& \ j \ne 1
\nonumber\\
\tilde{ J}^{S-\frac{1}{2}} _{ii} =  {}^{S}C^{0}_{ii}- \frac{1}{\epsilon_i-\epsilon_1} + \Gamma^{\{\nu_1 \dots \nu_{\Omega-1}\}}_1(\epsilon_i) = J^{S-\frac{1}{2}} _{ii}  \ \ \ &\forall& \ i \ne 1
\nonumber\\
\tilde{J}^{S-\frac{1}{2}} _{ij}=  {}^{S}C^{0}_{ij}  = J^{S-\frac{1}{2}} _{ij} \ \ \ &\forall& \ i \ne 1\ ,\ j \ne i ,\nonumber\\
\eea

\noindent since
\bea
 {}^{S-\frac{1}{2}}C^n_{11} &=&\left[\displaystyle\sum_{E\in S^{(2S-1-n)}} \frac{1}{\prod_{k=1}^{2S-1-n}(\epsilon_1 - E_k)}\right] =
  {}^{S}C^{n+1}_{11}
\nonumber\\
 {}^{S-\frac{1}{2}}C^n_{1j} &=&-\left[\displaystyle\sum_{p=n}^{(2S-1)-1}\displaystyle\sum_{E \in S^{(p-n)}_{\hat{j}}} \frac{2S-1-p}{(\epsilon_{1}- \epsilon_{j})^{2S-1-p}}\frac{1}{\prod_{k = 1}^{p-n}(\epsilon_{1} - E_{k})}\right]
\nonumber\\ 
 &=&-\left[\displaystyle\sum_{p'=n+1}^{2S-1}\displaystyle\sum_{E \in S^{(p'-n-1)}_{\hat{j}}} \frac{2S-p'}{(\epsilon_{1}- \epsilon_{j})^{2S-p'}}\frac{1}{\prod_{k = 1}^{p'-n-1}(\epsilon_{1} - E_{k})}\right] =  {}^{S}C^{n+1}_{1j}  \ \ \ \forall j \ne 1
\nonumber\\ 
{}^{S-\frac{1}{2}}C^0_{ii} &=&  \left[ \frac{2S-1}{\epsilon_i-\epsilon_1} + \displaystyle\sum_{k \ne i \ne 1}^{N}\frac{1}{\epsilon_i-\epsilon_k} \right] = {}^{S}C^{0}_{ii} -  \frac{1}{\epsilon_i-\epsilon_1}  \ \ \  \forall i \ne 1
 \eea

This shows that the first recursive condition concerning the value of the residues at $\nu_i = \epsilon_1$ is met by the determinant of matrix (\ref{finalJ}).

With the same logic,  using
\bea
\lim_{\nu_{\Omega} \to \epsilon_N} \Gamma_n^{{\{\nu_1 \dots \nu_\Omega\}}}(\epsilon_j) = \Gamma_n^{\{\nu_1 \dots \nu_{\Omega-1}\}}(\epsilon_j) + \frac{1}{\epsilon_{N}-\epsilon_j} \Gamma_{n-1}^{\{\nu_1 \dots \nu_{\Omega-1}\}}(\epsilon_j) \ \forall  j \ne N
\nonumber\\ 
\mathrm{Res}_{\nu_{\Omega} = \epsilon_N} \Gamma_n^{\{\nu_1 \dots \nu_\Omega\}}(\epsilon_N) = \Gamma_{n-1}^{\{\nu_1 \dots \nu_{\Omega-1}\}}(\epsilon_N) , 
\eea
\noindent we find
\bea
\mathrm{Res}_{\nu_{\Omega} = \epsilon_N}\det J^S = \det \tilde{J}^{\hat{N}} ,
\eea
\noindent with
\bea
\tilde{J}^{\hat{N}}_{NN} = 1
\nonumber\\ 
\tilde{J}^{\hat{N}}_{Nj} = 0 \ \ &\forall& \ j \ne N
\nonumber\\
\tilde{J}^{\hat{N}}_{ii} =  {}^SC^{0}_{ii} - \frac{1}{{\epsilon_i-\epsilon_N}} +  \Gamma^{\{\nu_1 \dots \nu_{\Omega-1}\}}_1(\epsilon_i) =J^{\hat{N}}_{ii}   \ \ &\forall &\ i \ne 1 \ne N
\nonumber\\
\tilde{J}^{\hat{N}}_{ij} = {}^SC^{0}_{ij}  = J^{\hat{N}}_{ij} \ \ &\forall& \ i \ne 1 \ne N \ ,\ j \ne i
\nonumber
\eea
\bea
\tilde{J}^{\hat{N}}_{1j} = (2S)!\left({}^SC^{2S-1}_{1j} \Gamma^{\{\nu_1 \dots \nu_{\Omega-1}\}}_{2S-1}(\epsilon_1) \right.\nonumber\\ \left. \ \ + \sum_{n = 0}^{2S-2}\left[\frac{{}^SC^{n+1}_{1j}}{\epsilon_N-\epsilon_1}  + {}^SC^{n}_{1j}\right]\Gamma^{\{\nu_1 \dots \nu_{\Omega-1}\}}_n(\epsilon_1)\right) =  J^{\hat{N}}_{1j} \ \ &\forall& \ j \ne 1 
\nonumber\\
\nonumber\\
\tilde{J}^{\hat{N}}_{11} =(2S)!\left( {}^SC^{2S}_{11}\Gamma^{\{\nu_1 \dots \nu_{\Omega-1}\}}_{2S}(\epsilon_1) \right.\nonumber\\ \ \ \left.+ \sum_{n = 0}^{2S-1}\left[\frac{{}^SC^{n+1}_{11}}{\epsilon_N-\epsilon_1} + {}^SC^{n}_{11}\right] \Gamma^{\{\nu_1 \dots \nu_{\Omega-1}\}}_n(\epsilon_1)\right)  =  J^{\hat{N}}_{11}, 
\eea

\noindent where  $J^{\hat{N}}$ is the $(N-1) \times (N-1)$ matrix which corresponds to the matrix one would obtain from (\ref{finalJ}) for  a system made out of a spin $S$ and a collection of $N-2$ spins $\frac{1}{2}$ labelled from $2$ to $N-1$. Since line $N$ is turned into $ ( 0 \ 0 \ \dots \ 0 \ 1 )$, the resulting determinant giving the residue is then the one of the $(N-1) \times (N-1)$ matrix $J^{\hat{N}}$. Indeed, using the notation $ {}^{S; \hat{N}}C^{n}_{ij} $ for the coefficients appearing in $J^{\hat{N}}$, we do have :

\bea
 {}^{S; \hat{N}}C^{0}_{ii} &=& \left[ \frac{2S}{\epsilon_i-\epsilon_1} + \displaystyle\sum_{k \ne i \ne 1}^{N-1}\frac{1}{\epsilon_i-\epsilon_k} \right] \nonumber\\ &=&
 \left[ \frac{2S}{\epsilon_i-\epsilon_1} + \displaystyle\sum_{k \ne i \ne 1}^{N}\frac{1}{\epsilon_i-\epsilon_k} \right] 
 - \frac{1}{{\epsilon_i-\epsilon_N}} = {}^SC^{0}_{ii} - \frac{1}{{\epsilon_i-\epsilon_N}} \ \ \forall i \ne 1,N
 \\
 {}^{S; \hat{N}}C^{n}_{1j} &=& \left[\displaystyle\sum_{E\in S^{(2S-n)}_{\hat{N}}} \frac{1}{\prod_{k=1}^{2S-n}(\epsilon_1 - E_k)}\right] \nonumber\\ &=&
\displaystyle\sum_{E\in S^{(2S-n)}} \frac{1}{\prod_{k=1}^{2S-n}(\epsilon_1 - E_k)}
-\frac{1}{(\epsilon_1 - \epsilon_N)} \left(\sum_{E\in S^{(2S-n-1)}} \frac{1}{\prod_{k=1}^{2S-n-1}(\epsilon_1 - E_k)}\right)\nonumber\\
&=& {}^SC^{n}_{1j} + \frac{1}{\epsilon_N-\epsilon_1} {}^SC^{n+1}_{1j},
\eea

\noindent where the sum over $S^{(2S-n)}_{\hat{N}}$ (which excludes $\epsilon_N$) has been rewritten as the sum over $S^{(2S-n)}$
(now including $\epsilon_N$) from which we remove every term which contains at least one instance of $\epsilon_N$, terms which are built from multiplying $\frac{1}{(\epsilon_1 - \epsilon_N)}$ by the terms of cardinality $n-1$. The only exceptions to this last rule are the coefficients ${}^{S; \hat{N}}C^{2S}_{11} = 1 = {}^{S}C^{2S}_{11} $ and ${}^{S; \hat{N}}C^{2S-1}_{1j(\ne 1,N)} = \frac{1}{\epsilon_j -\epsilon_1} = {}^{S}C^{2S-1}_{1j(\ne 1,N)}$ showing the complete equivalence of the determinants of $\tilde{J}^{\hat{N}}$ and $J^{\hat{N}}$.
 
Having shown that every residue at $\epsilon_1$ of the determinant of $J^S$ gives back the similar determinant of a matrix $J^{S-\frac{1}{2}}$ (multiplied by $2S$) and that residues at a different $\epsilon_j$ gives the determinant of the matrix $J^{\hat{j}}$ with spin $j$ removed, we know that to complete the proof we now simply need to show that the limit when every $\nu_i \to \infty$ of the proposed determinant is going to zero. Within the determinant, the different terms in $\Gamma_{n > 0}(\epsilon_i)$ cancel in this limit, so that only the determinant of the matrix containing the constant coefficients remains:

\bea
J_{ij}^{lim}= 
\left\{
\begin{array}{cc}
(2S)!\displaystyle\sum_{E\in S^{(2S)}} \frac{1}{\prod_{k=1}^{2S}(\epsilon_1 - E_k)}   &   i=j=1  \\
-(2S)!\displaystyle\sum_{n=0}^{2S-1}\displaystyle\sum_{E \in S^{(n)}_{\hat{j}}} \frac{2S-n}{(\epsilon_{1}- \epsilon_{j})^{2S-n}}\frac{1}{\prod_{k = 1}^{n}(\epsilon_{1} - E_{k})}  & i=1 \ne j\\
 \frac{2S}{\epsilon_i-\epsilon_1} + \displaystyle\sum_{k \ne i \ne 1}^{N}\frac{1}{\epsilon_i-\epsilon_k} &  i = j \ne 1\\
 \frac{1}{\epsilon_j-\epsilon_i} &  i \ne j \ne 1
 
  \end{array}
\right. .
\label{detlim}
\eea

By adding all the matrix elements $J^{lim}_{i,(j\ne1)}$ of a given line to the first column, the determinant remains unchanged while the elements of the first column become:

\bea
J_{i1}^{lim}= 
\left\{
\begin{array}{cc}
\left(\displaystyle\sum_{E\in S^{(2S)}} \frac{(2S)!}{\prod_{k=1}^{2S}(\epsilon_1 - E_k)} - \displaystyle\sum_{j = 2}^{N}\displaystyle\sum_{n=0}^{2S-1}\displaystyle\sum_{E \in S^{(n)}_{\hat{j}}} \frac{2S-n}{(\epsilon_{1}- \epsilon_{j})^{2S-n}}\frac{(2S)!}{\prod_{k = 1}^{n}(\epsilon_{1} - E_{k})}\right) &   i=1  \\
 \frac{1-2S}{\epsilon_1-\epsilon_i} &  i \ne 1
  \end{array}
\right. .
\eea

\noindent Considering a given element of $\displaystyle\sum_{E\in S^{(2S)}} \frac{1}{\prod_{k=1}^{2S}(\epsilon_1 - E_k)}$ for which each factor of the type $\frac{1}{(\epsilon_{1}-\epsilon_i)}$ appears with a given power between $0$ and $2S$ and can be written as $\displaystyle\prod_{i=2}^{N}\frac{1}{(\epsilon_1-\epsilon_i)^{x_i}}$ with $0\leq x_i \leq 2S$. In $\displaystyle\sum_{j = 2}^{N}\displaystyle\sum_{n=0}^{2S-1}\displaystyle\sum_{E \in S^{(n)}_{\hat{j}}} \frac{2S-n}{(\epsilon_{1}- \epsilon_{j})^{2S-n}}\frac{1}{\prod_{k = 1}^{n}(\epsilon_{1} - E_{k})}$, $\displaystyle\prod_{i=2}^{N}\frac{1}{(\epsilon_1-\epsilon_i)^{x_i}}$ appears $x_i$ times because of this factor: $\frac{2S-n}{(\epsilon_{1}- \epsilon_{j})^{2S-n}}$. By completing the sum over $j$ and knowing that $\displaystyle\sum_{i=2}^{N}x_i = 2S$, each of the $\displaystyle\prod_{i=2}^{N}\frac{1}{(\epsilon_1-\epsilon_i)^{x_i}}$ appears $1-2S$ times in $J_{11}$. Therefore, all the matrix elements in the first column became :

\bea
J_{i1}^{lim}= 
\left\{
\begin{array}{cc}(2S)! \left(\displaystyle\sum_{E\in S^{(2S)}} \frac{1-2S}{\prod_{k=1}^{2S}(\epsilon_1 - E_k)}\right)  &   i=1  \\
 \frac{1-2S}{\epsilon_1-\epsilon_i} &  i \ne 1
 
  \end{array}
\right. ,
\eea

\noindent while the other columns have been left untouched. Comparing these new elements with the previous ones from equation (\ref{detlim}):

\bea
J_{i1}^{lim}= 
\left\{
\begin{array}{cc}
(2S)! \left(\displaystyle\sum_{E\in S^{(2S)}} \frac{1}{\prod_{k=1}^{2S}(\epsilon_1 - E_k)}\right)  &   i=1  \\
 \frac{1}{\epsilon_1-\epsilon_i} &  i \ne 1
 
  \end{array}
\right. ,
\eea

\noindent we are left with $\det J^{lim} = (1-2S)\det J^{lim} $. This unambiguously implies that $\det J^{lim}  = 0$.

Provided the proposed expression is valid for a collection of $N$ (and $N-1$) spins $\frac{1}{2}$, we showed that $Z^{S=1}$ is then indeed given by $\mathrm{det} J^{S=1}$. Recursively we therefore proved the validity of $Z^{S=3/2}=\mathrm{det} J^{S=3/2}$ and so on, for an arbitrary value of $S$.

For $S=1/2$, eq. (\ref{finalJ}) is given explicitly by:
\bea
J_{ii} &=& \displaystyle\sum_{n=0}^{1}\left[\displaystyle\sum_{E\in S^{(1-n)}} \frac{1}{\prod_{k=1}^{1-n}(\epsilon_i - E_k)}\right]\Gamma_n(\epsilon_i) \ \ \forall i
\nonumber\\
J_{ij} &=& \displaystyle\sum_{n=0}^{1-1}\left[\displaystyle\sum_{p=n}^{1-1}\displaystyle\sum_{E \in S^{(p-n)}_{\hat{j}}} \frac{1-p}{(\epsilon_{i}- \epsilon_{j})^{1-p}}\frac{1}{\prod_{k = 1}^{p-n}(\epsilon_{1} - E_{k})}\right]\Gamma_n(\epsilon_1)  \ \ \forall i \ne j 
\nonumber\\ &=&  \frac{1}{(\epsilon_{i}- \epsilon_{j})}
\eea 
which is indeed equal to eq. (\ref{Jwithsigns}), therefore completing the proof.
 
\section{Conclusions}

The scalar product of an arbitrary dual state and an arbitrary normal state is obviously writable as the domain wall boundary partition function for which a determinant expression was derived in this work.

Since a system of spins of finite length allows the construction of a dual Bethe Ansatz obtained from a straightforward application of the quantum inverse scattering method, any eigenstate of the system $\prod_{i=1}^{M} S^{+}(\lambda_i) \left|\Downarrow_S, \downarrow \dots \downarrow\right>$ is in one to one correspondence with a dual representation $\prod_{i=1}^{\Omega-M} S^{-}(\mu_i) \left|\Uparrow_S, \uparrow \dots \uparrow\right>$, this also gives a direct access to the projection of an arbitrary off-the-shell Bethe state on the (dual) eigenbasis of any such integrable Hamiltonian defined with an arbitrary value of $g$. This form the starting point of quench-like problems, namely the decomposition of an initial condition onto the eigenbasis of the Hamiltonian which controls the subsequent time-evolution.

Know that, in a similar fashion to the work carried out in \cite{faridet,bosonhugo}, the expressions found here could also be reused to define form factors of local spin operators. Moreover, the approach used in this work should be generalizable to models built out of a collection of  spins where each spin $i$ has its own arbitrary length $S_i$. We choose, however, to defer these issues to subsequent publications.

\end{document}